\documentclass[12pt,preprint]{aastex}

\begin{document}
\title{Lowering the Characteristic Mass of Cluster Stars  
by Magnetic Fields and Outflow Feedback} 
\shorttitle{{\sc Characteristic Stellar Mass}}
\author{ Zhi-Yun Li$\!$\altaffilmark{1}, Peng Wang$\!$\altaffilmark{2},
Tom Abel$\!$\altaffilmark{2} and Fumitaka   Nakamura$\!$\altaffilmark{3} }
\altaffiltext{1}{Astronomy Department, P.O. Box 400325, 
University of Virginia, Charlottesville, VA 22904; zl4h@virginia.edu}
\altaffiltext{2}{KIPAC, SLAC, and Physics Department, Stanford
  University, Menlo Park, CA 94025} 
\altaffiltext{3} {National Astronomical Observatory, Mikata, Tokyo
  181-8588, Japan, and visiting astronomer at Institute of Space and 
Astronautical Science, Japan Aerospace Exploration Agency, 
3-1-1 Yoshinodai, Sagamihara, Kanagawa 229-8510, Japan}

\begin{abstract}
Magnetic fields are generally expected to increase the 
characteristic mass of stars formed in stellar clusters, 
because they tend to increase the effective Jeans mass. We 
test this expectation using adaptive mesh refinement (AMR) 
magnetohydrodynamic simulations of cluster formation in 
turbulent magnetized clumps of molecular clouds, treating 
stars as accreting sink particles. We find that, contrary 
to the common expectation, a magnetic field of strength in 
the observed range decreases, rather than increases, the 
characteristic stellar mass. It (1) reduces the number of 
intermediate-mass stars that are formed through direct 
turbulent compression, because sub-regions of the clump 
with masses comparable to those of stars are typically 
magnetically subcritical and cannot be compressed directly 
into collapse, and (2) 
increases the number of low-mass stars that are produced 
from the fragmentation of dense filaments. The filaments 
result from mass accumulation along the field lines. 
In order to become magnetically supercritical and fragment, 
the filament must accumulate a large enough column density 
(proportional to the field strength), which yields a high 
volume density (and thus a small thermal Jeans mass) that 
is conducive to forming low-mass stars. We 
find, in addition, that the characteristic stellar mass 
is reduced further by outflow feedback. The conclusion 
is that both magnetic fields and outflow feedback are 
important in shaping the stellar initial mass function 
(IMF).    
\end{abstract}

\section{Introduction}
\label{intro} 

How the initial mass function (IMF) of stars originates is one of the
most basic questions that a complete theory of star formation must 
answer. It is also one of the most difficult (see reviews by Bonnell 
et al. 2007 and McKee \& Ostriker 2007 and references therein). Many 
ideas have been advanced to explain the IMF, including Zinnecker 
(1982), Adams \& Fatuzzo (1996), Elmegreen (1997), Padoan 
\& Nordlund (2002),  Larson (2003), Shu et 
al. (2004), Hennebelle \& Chabrier (2008), and Kunz \& Mouschovias 
(2009), among others. Particularly intriguing is the proposal 
by Padoan \& Nordlund (2002) that the 
IMF is essentially determined by the supersonic turbulence, which 
controls the density distribution in the highly inhomogeneous 
clouds. They used the Jeans criterion to determine analytically 
the mass distribution of the gravitationally unstable regions, 
which was taken to represent the stellar IMF. Both ingredients 
of the theory, the turbulent 
structuring of cloud density and the criterion for gravitational 
collapse, are strongly affected, however, by a dynamically important 
magnetic field. Some magnetic effects on the IMF, 
such as the magnetic cushion of turbulent compression (Padoan \& 
Nordlund 2002) and the additional cloud support by 
magnetic pressure 
(Hennebelle \& Chabrier 2008), can be incorporated into the 
analytic theory approximately. Other equally, if not more, important 
magnetic effects are not captured by the theory, and will be the 
focus of our investigation. 

Our study will concentrate on the parsec-scale, cluster-forming 
dense clumps of molecular clouds, where the majority of stars, 
especially massive stars, are thought to originate (Lada \& Lada 
2003), and where magnetic fields have been observed in some 
cases. For example, in the nearest region of active massive star 
formation, OMC1, a well-ordered field is inferred from polarized 
dust emission (e.g., Vaillancourt et al. 2008). Crutcher et al. 
(1999) obtained, from CN Zeeman 
measurements, a line-of-sight 
field strength $B_{\rm los}=360$~$\mu$G for this region, which 
corresponds to a dimensionless mass-to-flux ratio $\lambda = 2\pi G^{1/2} 
M/\Phi\sim 4.5$. 
%, 
%based on the measured $B_{\rm los}$ and an estimate of the 
%column density along the same line of sight. 
% based on CN abundance of CN/H2=4x10^-9??
The ratio is close to the median value $\lambda\sim 6$ obtained by
Falgarone et al. (2008) for a sample of dense clumps of massive 
star formation. Correcting for the projection effects may reduce the 
ratio by a factor of 2-3 (Shu et al. 1999), yielding a median value 
for the intrinsic mass-to-flux ratio of 2-3. 

A magnetic field corresponding to $\lambda \sim$a few would not be 
strong enough to support the clump against 
gravitational collapse by itself. It can, however, change the 
mass distribution of the stars formed. This is 
because individual stars form out of sub-regions of the clump, 
which are generally more strongly magnetized (relative to their 
masses) than the clump as a whole (i.e., with smaller values 
of $\lambda$, Tilley \& Pudritz 2007; Dib et al. 2007), because 
the mass of a region (which is proportional to volume) decreases 
faster with its size than the magnetic flux (which is proportional 
to area). In particular, sub-regions smaller than the clump by a 
factor comparable to $\lambda$ are expected to be 
magnetically subcritical (with $\lambda < 1$) in general. It 
would be hard to directly compress such regions into collapse by 
turbulence in the ideal MHD limit (see, however, Nakamura \& Li 
2005 and Basu et al. 2009 for studies that include ambipolar 
diffusion). Another effect is that the magnetic forces act on 
the turbulent flows anisotropically, channeling matter along 
the field lines into dense, flattened structures that can 
subsequently fragment thermally. 

In this Letter, we seek to quantify the above effects using 
the cluster formation simulations of Wang et al. (2010), 
which were carried out with 
an AMR MHD code that includes sink particles and outflow 
feedback. We find, in \S~\ref{numerical}, that the characteristic 
mass of the IMF is lowered by both the magnetic field and 
outflow feedback. The lowering of the characteristic stellar 
mass by magnetic fields is somewhat counter-intuitive. 
It is interpreted physically using a so-called ``magnetically 
critical Jeans mass'' (eq.~[\ref{MagneticallyCriticalJeansMass}]) 
in \S~\ref{discussion}. 

\section{Result}
\label{numerical}

In Wang et al. (2010), we simulated cluster formation in moderately 
condensed, isothermal ($T=20$~K), clump of 2~pc in size 
and $1641$~M$_\odot$ in mass. We added to the clump, one by one, an 
initial turbulent velocity field of rms Mach number 9 (Model HD), 
an initially uniform magnetic field of $100$~$\mu$G (corresponding 
to a dimensionless mass-to-flux 
ratio of 1.4 for the clump as a whole 
and a larger value of 3.3 for the central, 
high-density part; Model MHD), and mechanical feedback from 
collimated outflows driven by accreting stellar objects (Model 
WIND). For simulation details we refer the reader to Wang et al. 
(2010), where we 
also presented results on the global star formation rate 
and especially the formation of the most massive object in 
each simulation. Here, we will concentrate on the 
distribution of the stellar masses, with an emphasis on low 
and intermediate mass objects. 

Figure~1 shows the stellar mass distributions for all three 
models. It is a log-linear plot of the number of stars in 
logarithmic mass bins of $\Delta [{\rm log}(M)]=0.2$ (where $M$ 
is the stellar mass in M$_\odot$). The 
distributions are all evaluated at the time when $\sim 16\%$ of 
the total mass has been converted into stars. It corresponds 
to 1.16~Myr for Model WIND (the last output), and 0.576 and 0.723~Myr
for Model HD and MHD, respectively; for comparison, the 
free-fall time is $0.210$~Myr at the clump center initially 
and $0.475$~Myr in a sphere of 1~pc in radius. 
% containing 1216 solar masses; t_ff=0.565 Myr for the whole square box

\begin{figure}
%\plottwo{TotalMass.eps}{TotalAccretionRate.eps}
\epsscale{1.0}
\plotone{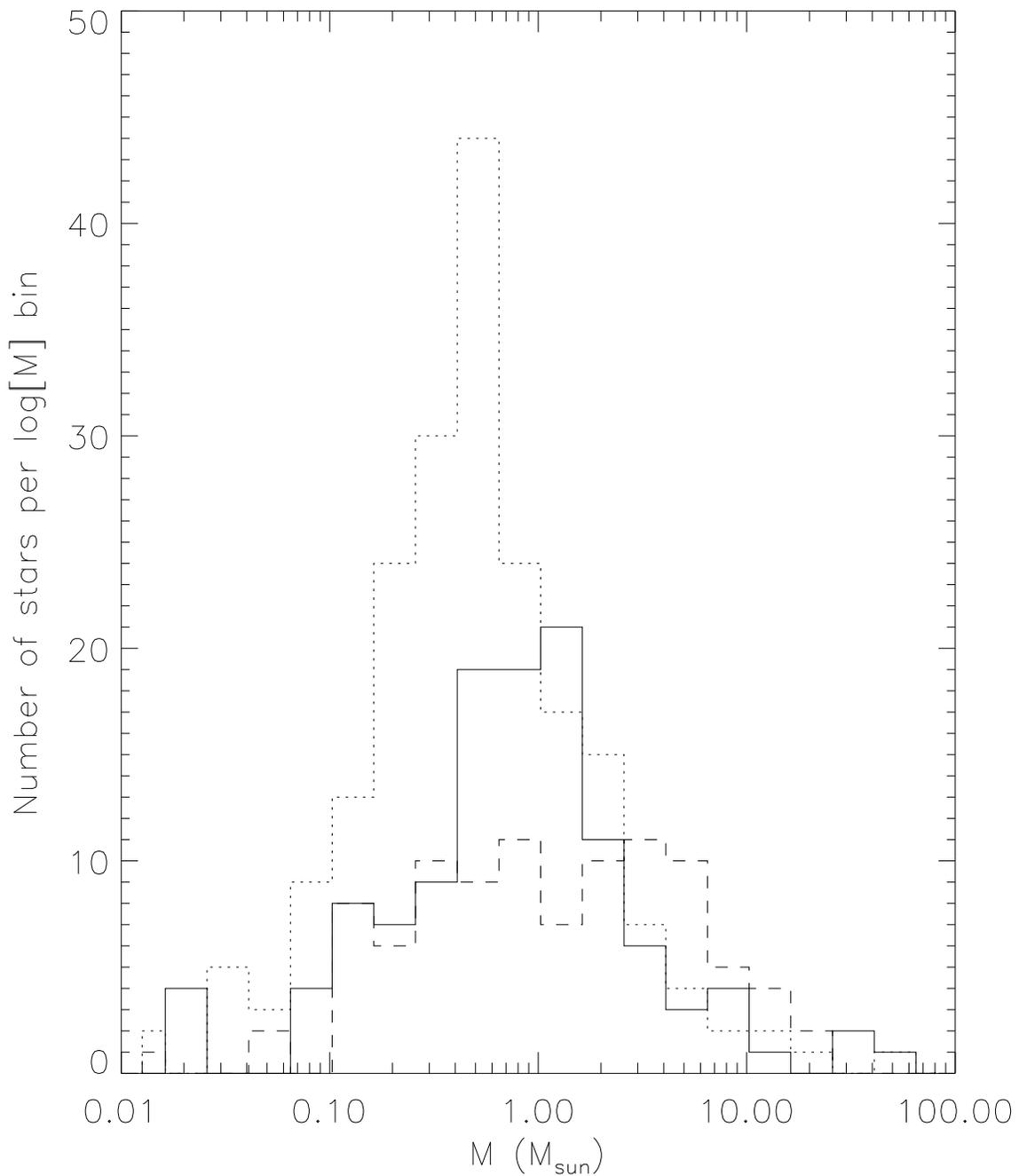}
\caption{The number of stars in logarithmic mass bins of $\Delta [{\rm 
log}(M)]=0.2$ for Model HD 
(dashed), MHD (solid) and WIND (dotted), when $16\%$ of the clump 
mass has been converted into stars.} 
\label{imf}
\end{figure}

Before proceeding to analyze the numerically obtained mass
distributions in Fig.~1, we should caution the reader that 
the mass of a star can be affected by many factors that are 
only crudely modeled in the current generation of large-scale 
cluster formation simulations, including our own. These include 
(1) outflow feedback, which can potentially remove a large 
fraction, perhaps the majority of the mass of a star-forming 
core (Shu et al. 1987; Matzner \& McKee 2000; Myers 2008), (2) 
magnetic fields, which affect not only how dense cores form but 
also how they collapse, and (3) radiative feedback which, once 
a protostellar system (protostar plus disk) appears, heats up 
the circumstellar material and suppresses fragmentation close 
to the central object (e.g., Krumholz et al. 2007; 
Offner et al. 2009; Price \& Bate 2009; 
Smith et al. 2009; Urban et al. 2009). It is currently not feasible 
to treat simultaneously all these factors in detail. Our strategy 
is to focus on magnetic fields and outflow feedback, which are 
less well studied, and treat the effects of radiative feedback 
as simply as possible. 

Specifically, we crudely capture the suppression of fragmentation 
near a star due to radiative feedback using a sink-particle merging 
algorithm, following Krumholz et al. (2004, see also
discussion in Federrath et al. 2010). It  
eliminates very low mass ($ < M_{\rm merg}=0.01$~M$_\odot$) 
sink particles within a distance 
$l_{\rm merg} = 1000$~AU of an existing star (with mass 
$ > M_{\rm merg}$). 
Experimentation shows that increasing $M_{\rm merg}$ to 
$0.1$~M$_\odot$ does not change the star formation rate or
the stellar mass distribution significantly. The adopted 
merging distance $l_{\rm merg}$ is comparable to the 
radius of the so-called ``sphere of thermal influence'' 
(where the temperature of the radiatively heated 
protostellar envelope drops to the ambient 
value) of an object of solar mass and luminosity (see 
equation~[14] and Figure~2 of Adams \& Shu 1985).  
We find that increasing $l_{\rm merg}$ from 1000~AU to 
2000~AU has little effect on the star formation rate and
stellar mass distribution. Decreasing it to $600$~AU 
(or three times the size of the finest cell) does not
change the total star formation rate, but can change 
the number and masses of stars significantly, by up to 
50\%. We believe, however, that the smaller merging 
distance is less realistic, since the gas on the 
smaller scale is expected to be more strongly heated.
The larger merging distance prevents a collapsing core 
from breaking up into many small pieces, in agreement 
with the fact that the best observed cores in nearby 
star-forming regions typically harbor one, at most a 
few, stellar systems (e.g., Mundy et al. 2000). 
Nevertheless, in view of our crude representation of 
radiative feedback, we elect to focus on the 
difference between the stellar mass distributions of 
the three simulations, which use the same sink particle 
treatment, rather than the distributions themselves.

\begin{figure}
\epsscale{0.75}
\plotone{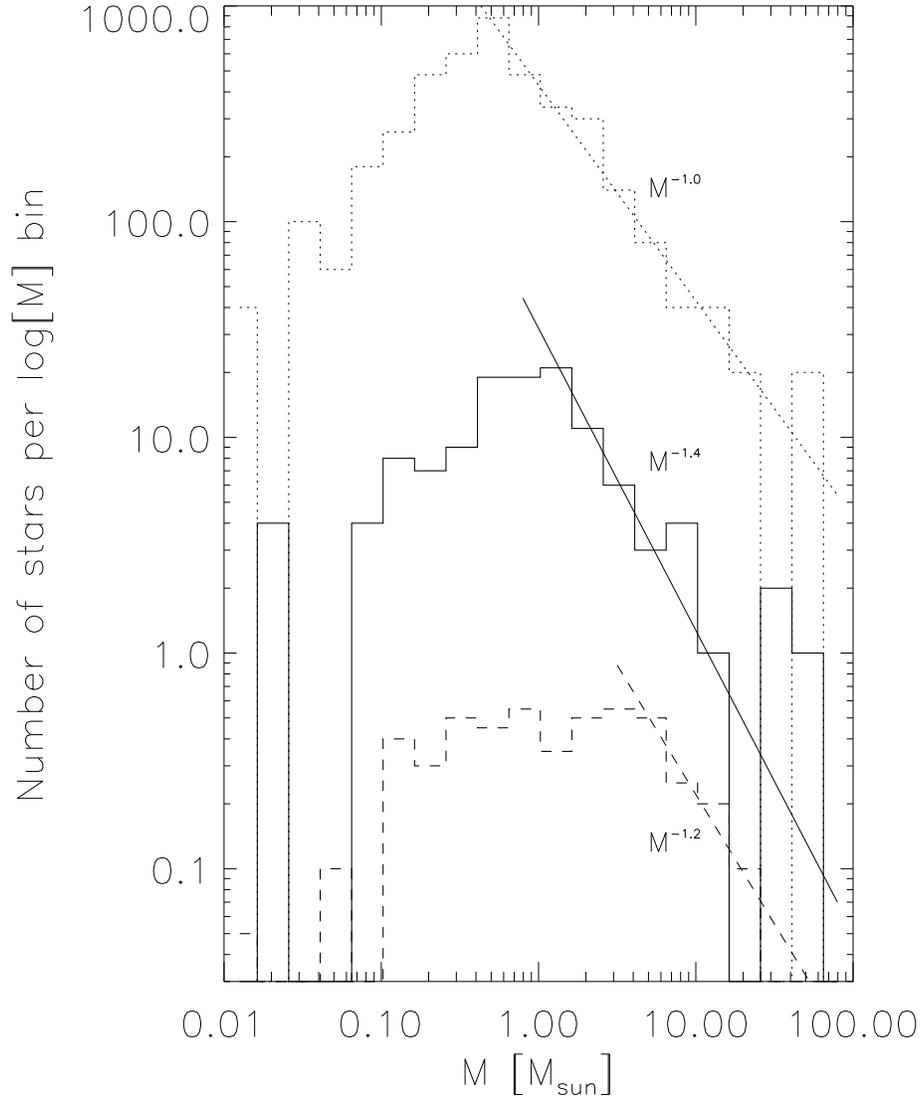}
\caption{Power-law fits to the high mass end of the stellar mass 
distributions of the three models: HD (dashed lines), MHD (solid) 
and WIND (dotted). The top (bottom) curve is raised (lowered) by 
a factor of 20 for clarity. } 
\label{imf_log}
\end{figure}

The magnetic field is solely responsible for the difference 
between Model HD and MHD. From Fig.~1, we see that its 
main effects are to reduce the number of those 
intermediate-mass stars with masses around $\sim 4$~M$_\odot$, 
and to increase the number of lower mass stars with masses 
around $\sim 1$~M$_\odot$. To be more quantitative, we 
fit the high mass end of the distribution with 
a power-law, $dN/d\log(M) \propto M^\Gamma$, as shown in 
Fig.~\ref{imf_log}. In all three cases the mass 
distribution deviates sharply from the power-law fit 
below a characteristic mass, $M_{\rm ch}$ (sometimes 
dubbed the ``knee'' of the IMF, e.g., Bonnell et al. 2006).  
The values of $M_{\rm ch}$ and power-law index $\Gamma$ 
are listed in Table~1. Note that the characteristic 
mass of $\sim 5~M_\odot$ in Model HD is not far from the 
initial Jeans mass near the clump center, $\sim 9.5~M_\odot$, 
consistent with Bonnell et al. (2006). The power-indexes 
are not well determined because of the limited number of 
stars in the power-law regime. Nevertheless, they are all 
comparable to the Salpeter value within uncertainties. The
characteristic masses are, however, significantly different. 
In particular, the magnetic field in Model MHD has apparently 
lowered the characteristic mass by a factor of $\sim 4$ 
compared to that in Model HD.      

\begin{figure}
\epsscale{1.0}
\plotone{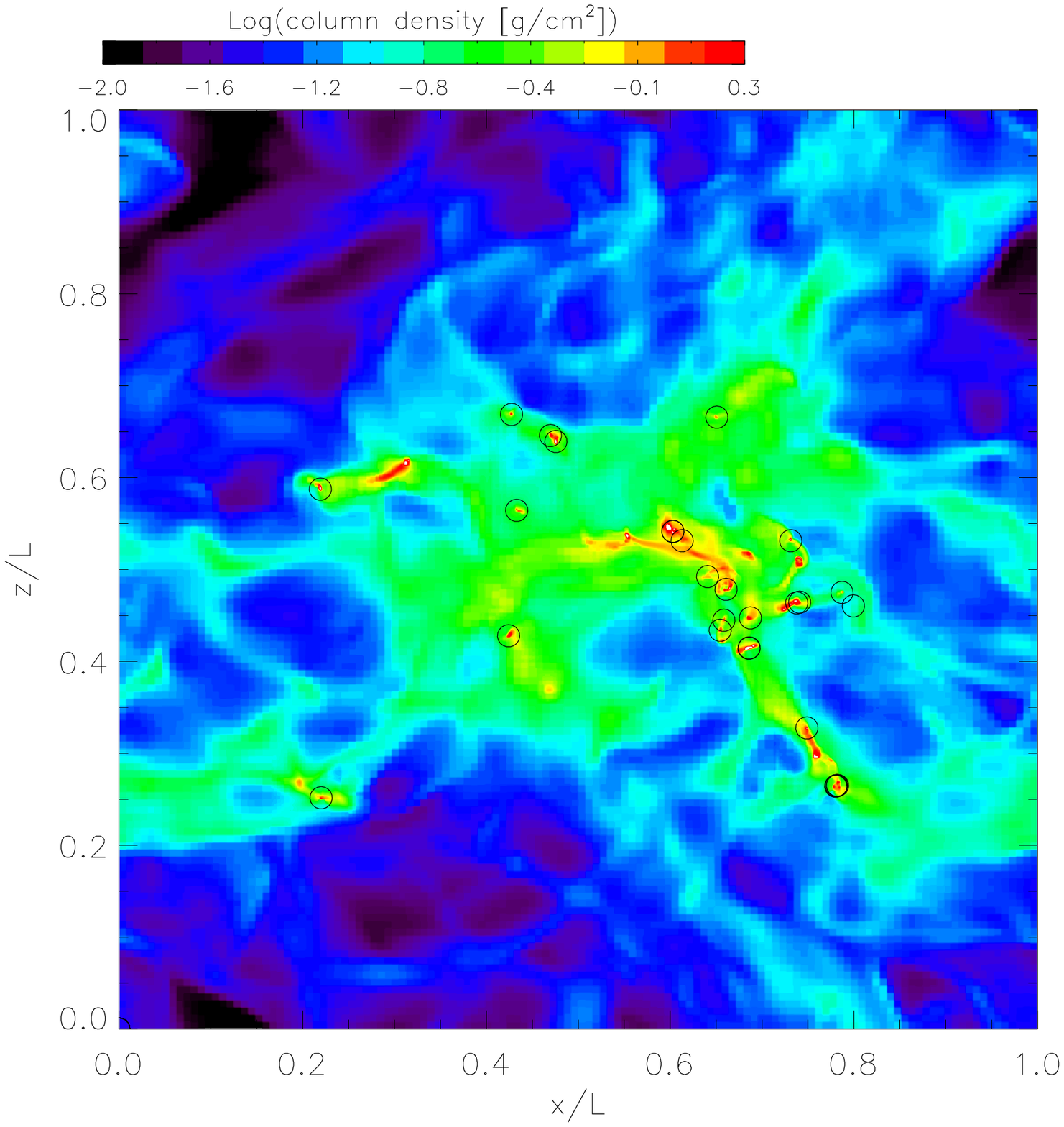}
\caption{Locations of the so-called ``HD excess stars'' (circled) at a 
representative time $t=0.498$~Myr on a column density map. 
The box size is $L=2$~pc.} 
\label{ExcessStars}
\end{figure}

To help understand why the field hampers the formation of 
intermediate-mass stars, we show 
in Figure~\ref{ExcessStars} a snapshot of the so-called ``HD excess 
stars,'' defined somewhat arbitrarily as those stars in HD model 
in Figs.~1 and 2 with masses in the three mass bins between 
$10^{1.2}$ and $10^{1.8}$~M$_\odot$. The majority of these 
stars are produced at relatively isolated regions, by localized 
converging flows in the initial turbulence. 
Such regions are largely absent in the MHD model, where a 
moderately strong global magnetic field is present. The 
reduction in the number of intermediate-mass stars makes the 
stellar mass distribution significantly narrower in the MHD 
model than in the HD model (see Table 1 and Figs.~1 and 2).

\begin{figure}
\epsscale{1.0}
\plotone{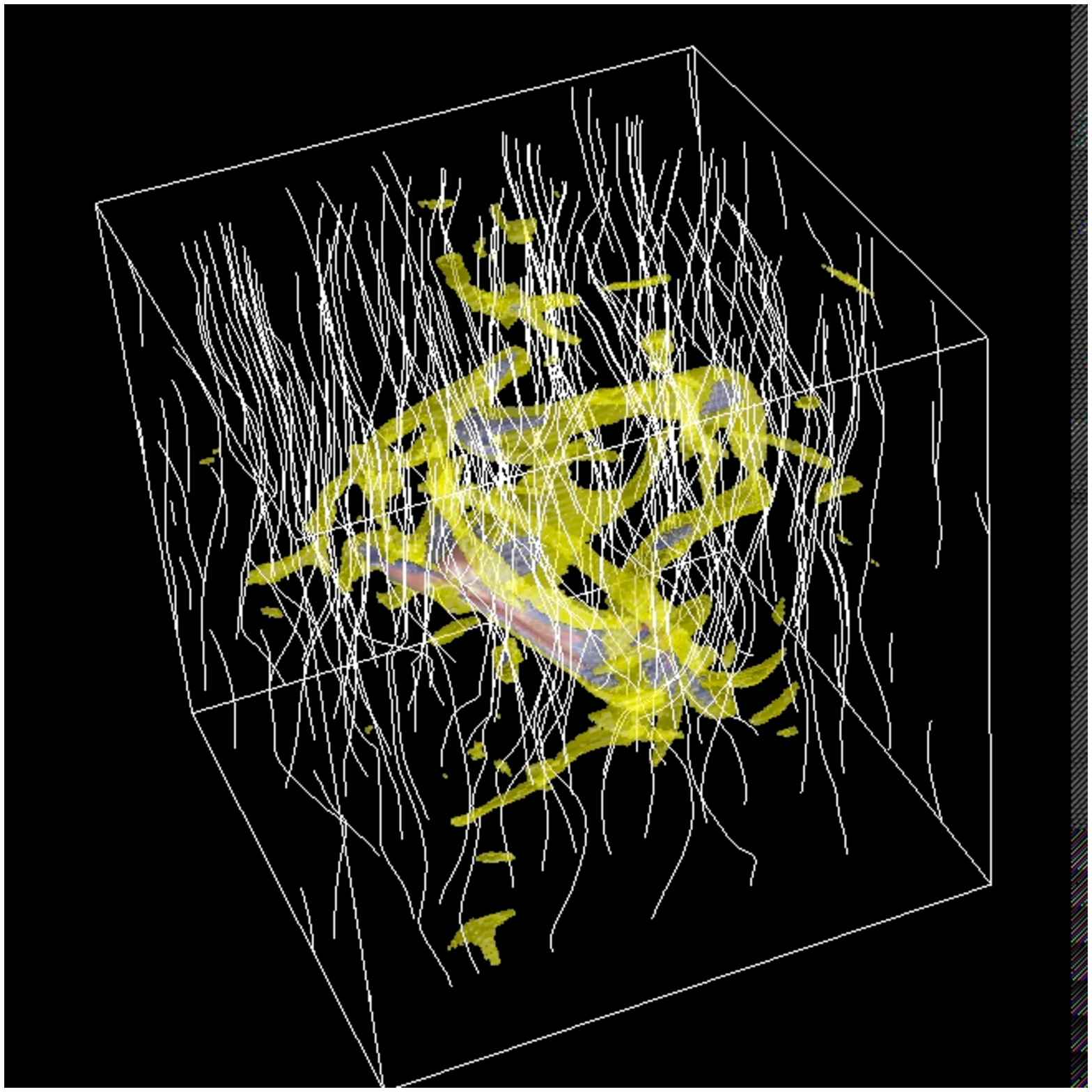}
\caption{3D view of a flattened filament in the MHD model. Plotted 
are isodensity surfaces and magnetic field lines (white) at a
relatively early time 0.262~Myr. The high density of the filament 
favors the formation of low-mass stars.} 
\label{MHD3D}
\end{figure}

The magnetic field in the MHD model interacts strongly with 
the turbulence. It changes the ``texture'' of the turbulent 
clump, by not only resisting the cross-field turbulent 
compression, but also promoting mass accumulation along the 
field lines into flattened sheets or filaments. Fig.~\ref{MHD3D} 
shows an example of the (flattened) filaments resulted  
from the intrinsically anisotropic magnetic forces. It 
is in such magnetically-induced filaments that the majority 
of the stars in the MHD model form. As we discuss in 
\S~\ref{discussion}, the high density in the filaments 
favors the formation of low-mass stars, which are more
numerous compared to the non-magnetic HD model.   

Outflow feedback is responsible for the difference between Model 
WIND and MHD. From Figs.~1 and 2, it is clear that the outflows 
have (1) shifted the characteristic mass $M_{\rm ch}$ to a 
lower value, from $\sim 1.3$~M$_\odot$ to $\sim 0.5$~M$_\odot$,
and (2) nearly doubled the number of stars, from 119 to 203 (
see Table~1). The lowering of the stellar mass by outflow 
feedback is perhaps not too surprising, 
because, individually, the outflow from a given star can remove 
part of the dense envelope that feeds the growing star (Shu et 
al. 1987; Matzner \& McKee 2000; Myers 2008) and, collectively, 
the feedback from all stars can prevent the clump
from rapid global collapse (Nakamura \& Li 2007; Wang et al. 
2010), lowering the total rate of stellar
mass accretion. There is, however, a large spread in both the 
rate and the duration of the mass accretion in stars of similar 
masses. The spread makes it difficult to develop a complete 
picture of how the outflows lower the characteristic stellar 
mass precisely. The number of stars produced per unit time 
(i.e., the stellar production rate) turns out to be insensitive 
to the outflow feedback. It takes longer, however, to convert 
the same amount of gas into stars in the presence of the 
feedback, which is the main reason for the larger number
of stars in Model Wind than in Model MHD.

\section{Discussion and Conclusion} 
\label{discussion}

We have shown that a magnetic field of strength in the observed range 
changes the IMF significantly, by impeding direct turbulent 
compression of relatively low density material 
into relatively massive stars and, more importantly, promoting the 
formation of dense, flattened structures that subsequently fragment 
preferentially into low-mass stars. 

The basic reason for the magnetic field to impede direct turbulent 
fragmentation is that a sub-region of the clump that is magnetically 
subcritical cannot be compressed into prompt collapse by turbulence. 
For a region of size $l$ in a clump of average field strength $B_0$ 
and mass density $\rho_0$, the characteristic dimensionless 
mass-to-flux ratio is 
\begin{equation}
\lambda_{l} = {2\pi G^{1/2} M_{l} \over \Phi_{l} }
\sim  { 2\pi G^{1/2} \rho_0 {l}^3 \over B_0 {l}^2}
=  { 2\pi G^{1/2} \rho_0 {l} \over B_0},  
\label{LocalLambda} 
\end{equation} 
which indicates that regions smaller than the characteristic size 
\begin{equation}
l_B = {B_0\over 2\pi G^{1/2} \rho_0} 
\label{CharacteristicSize}
\end{equation}
are typically magnetically subcritical (with $\lambda < 1$). The 
characteristic size is related to the clump size $L$ by $l_B 
\sim {L/\lambda_{\rm cp}}$ (where ``cp'' stands for ``clump''), 
since the mass-to-flux ratio of the clump as a whole is 
$
\lambda_{\rm cp} \sim { 2\pi G^{1/2} \rho_0 {L} /B_0},  
$
according to equation~(\ref{LocalLambda}). In other words, a typical
region that is smaller than the clump by a factor of $\sim \lambda_{\rm
  cp}$ cannot be compressed directly into collapse by
turbulence. The corresponding characteristic mass is
\begin{equation}
M_B \sim \rho_0 l_B^3 \sim {\rho_0 L^3\over \lambda_{\rm cp}^3} 
\sim {M_{\rm cp}\over \lambda_{\rm cp}^3} =  64 \left({M_{\rm 
cp}\over 10^3 M_\odot}\right)  
\left({2.5\over \lambda_{\rm cp}}\right)^3\ (M_\odot),
\label{AverageMagneticJeansMass} 
\end{equation}
where we have scaled the clump mass $M_{\rm cp}$ by $10^3$~M$_\odot$,  
and the clump mass-to-flux ratio by 2.5, comparable to the 
(projection-corrected) median value inferred from CN Zeeman 
observations (see discussion in \S~\ref{intro}). 

The characteristic mass $M_B$ is essentially the magnetic Jeans mass 
at the average clump density (see equation~[41] of McKee 1999, which 
gives a numerical value three times larger). It is to be compared 
with the average thermal Jeans mass
\begin{equation}
M_{\rm J} \approx 4.1 \left({L \over 1\ pc}\right)^{3/2}\left({10^3
   M_\odot \over M_{\rm cp} }\right)^{1/2} \left({T\over 10\ 
   K}\right)^{3/2}\ (M_\odot)
\label{JeansMass}   
\end{equation}
(e.g., Spitzer 1978). The average magnetic Jeans mass $M_B$ is 
therefore typically larger than the average thermal Jeans mass
$M_{\rm J}$, indicating that the formation 
of those stars with masses comparable to $M_{\rm J}$ 
or smaller by direct turbulent compression can be 
strongly hindered by the magnetic field. The most massive 
stars are expected to be least affected by the 
magnetic field. However, their 
formation may be governed more by the global clump collapse 
(retarded by outflow feedback and magnetic fields) than by 
direct turbulent compression (Wang et al. 2010; see also Smith 
et al. 2009). Indeed, massive ($> 10 M_\odot$) stars 
in Model MHD contain more mass than those in Model HD, 
indicating that their formation is not hampered by magnetic 
fields, unlike intermediate mass stars. 

The magnetic field can also promote fragmentation, because of the 
magnetic forces are intrinsically anisotropic. It is easier for 
the turbulent flows to move material along the field lines than 
across them. The net result is the formation of flattened, sheet-
or filament-like condensations, which are conducive to fragmentation.
Larson (1985) estimated the Jeans mass for a self-gravitating sheet
(thermally supported in the vertical direction) to be
\begin{equation}
M_{\rm J,sh} \approx {4.67\ a^4\over G^2 \Sigma}  
\label{JeansSheet}
\end{equation}
where $a$ is the isothermal sound speed, and $\Sigma$ the column 
density. In order to form stars, the sheet must first become 
magnetically supercritical, with a column density greater than 
the critical value $\Sigma_{\rm cr} = B_0/(2\pi G^{1/2})$. 
Substituting $\Sigma_{\rm cr}$ into equation~(\ref{JeansSheet}), 
we have
\begin{equation}
M_{\rm J, cr} \approx {9.34\ \pi\ a^4 \over G^{3/2} B_0} 
   = 0.78\  \left({L \over 1~pc}\right)^2 \left({10^3\ M_\odot \over
   M_{\rm cp} }\right) \left({ T\over 10~K}\right)^2 
     \left({\lambda_{\rm cp}\over 2.5}\right)\ (M_\odot) 
\label{MagneticallyCriticalJeansMass}
\end{equation}
which is the mass expected for stars formed through 
thermal fragmentation of magnetically critical filaments. A similar, 
but somewhat smaller, characteristic mass $M_0={\pi^2 a^4/(G^{3/2} 
B_0)}$ was defined in Shu et al. (2004) in their study of the 
collapse of magnetized singular isothermal spheres. The important 
conceptual point is that the magnetically critical Jeans mass 
$M_{\rm J, cr}$ (or $M_0$) decreases with an increasing magnetic 
field strength, because a stronger magnetic field can support 
a higher column density, which in turn leads to a higher volume density
(due to gravitational compression along the field lines) and
thus a lower thermal Jeans mass. It should be relatively 
insensitive to turbulence, which is expected to be weak 
at the relevant (high) densities. In any case, the magnetically 
critical Jeans mass is typically smaller than the average thermal 
Jeans mass, by a factor 
\begin{equation}
{M_{\rm J, cr}\over M_{\rm J}} \approx 0.19 \left({L \over 1~pc}
\right)^{1/2} \left({ 10^3 M_\odot \over  
M_{\rm cp}}\right)^{1/2} \left ({T\over 10~K}\right)^{1/2} 
\left({\lambda_{\rm cp} \over 2.5}
\right) 
\label{ratio} 
\end{equation}
which is relatively insensitive to clump parameters. It is therefore 
not too surprising that a moderately strong field can lower 
the characteristic mass of the stars formed in a cluster significantly, 
as we found numerically. Magnetism is another way of setting a 
stellar mass scale that is distinct from the one advocated by Larson 
(2005), based on detailed thermal physics of molecular cloud material 
(e.g., Jappsen et al. 2005 and Bonnell et al. 2006).

To summarize, we have demonstrated through both AMR MHD simulations 
(\S~\ref{numerical}) and analytic considerations 
(eqs.~[\ref{LocalLambda}]-[\ref{ratio}]) that moderately 
strong magnetic fields of 
the observed strengths (corresponding to a dimensionless mass-to-flux 
ratio of a few) are important in shaping the mass distribution 
of stars produced in turbulent, cluster-forming, dense clumps. 
The basic reason is that the field interacts strongly with the 
turbulent flows, which affects how the flows create regions 
unstable to gravitational collapse and star formation. 
The magnetic field
prevents sub-regions of the clump that are magnetically 
subcritical (with masses less than the average magnetic Jeans 
mass $M_B$) from being compressed directly into collapse by 
the turbulence. The magnetic resistance to turbulent compression 
has apparently reduced the number of intermediate mass stars 
formed in our simulations. More importantly, the 
intrinsically anisotropic 
magnetic forces channel part of the clump material along the 
field lines into dense structures, where the high density 
promotes the formation of low-mass stars with masses comparable 
to the magnetically critical Jeans mass $M_{\rm J,cr}$, 
which lowers the characteristic stellar mass. The mass is further 
reduced by outflow feedback, which 
affects not only the individual stars that drive the outflows, 
but also the dynamics of the clump as a whole. We conclude that 
magnetic fields and outflow feedback are important factors that 
should be accounted for in a complete theory of the stellar IMF.  
%
%Texture and anisotropy...
% 

\acknowledgments
This work was supported in part by NASA grants (NNG06GJ33G and 
NNX10AH30G) and a Grant-in-Aid for Scientific Research of Japan 
(20540228).

\begin{deluxetable}{lllll}
%\tabletypesize{\scriptsize}
%\rotate
\tablecolumns{5}
\tablecaption{Parameters of Stellar Mass Distribution \label{table:first}}
%\tablewidth{\columnwidth}
\tablehead{
\colhead{Model}     & \colhead{ IMF exponent $\Gamma$} & \colhead{
  Characteristic mass $M_{\rm ch}$} & \colhead{$\Delta^a$}  &
\colhead{Star number}
}
\startdata
HD   & $-1.2$  & $ 5.0~M_\odot$  & $ 0.8$  &  96  \\
MHD   & $ -1.4$  &   $ 1.3~M_\odot$ & $ 0.8$  & 119  \\
WIND   & $-1.0$  & $0.5~M_\odot$ & $ 1.8$  & 203  \\  
\enddata
\tablecomments{a) Full width of $\log (M)$ at half maximum of the 
stellar mass distribution. }
\end{deluxetable}

\pagebreak
%\begin{thebibliography}{}

\centerline{\bf References}

\setlength{\parindent}{0in}
%\bibitem{1}
Adams, F. C. \& Fatuzzo, M. 1996, ApJ, 464, 256 

%\bibitem{2}
Adams, F. C. \& Shu, F. H. 1985, ApJ, 296, 655

%\bibitem{3}
Basu, S., Ciolek, G. E., Dapp, W. B. \& Wurster, J. 2009, New
Astronomy, 14, 483

%\bibitem{4}
Bonnell, I. A., Clarke, C. \& Bate, M. 2006, MNRAS, 368, 1296
 
%\bibitem{5}
Bonnell, I. A., Larson, R. B. \& Zinnecker, H. 2007, in Protostars and Planets V, eds. B. Reipurth, D. Jewitt, and K. Keil (University of Arizona Press),  p149

%\bibitem{6}
Crutcher, R. M. et al. 1999, ApJ, 514, 121

%\bibitem{7}
Dib, S., Kim, J, et al. 2007, ApJ, 661, 262

%\bibitem{8}
Elmegreen, B. G. 1997, ApJ, 486, 944

%\bibitem{9}
Federrath, C., Banerjee, R., Clark, P. C. \& Klessen, R. S. 2010, ApJ,
713, 269

%\bibitem{10}
Falgarone, et al. 2008, AA, 487, 247

%\bibitem{11}
Hennebelle, P. \& Chabrier, G. 2008, ApJ, 684, 395

%\bibitem{12}
Jappsen, A.-K., Klessen, R. S., Larson, R. B. et al. 2005, AA, 435, 611

%\bibitem{13}
Krumholz, M. R., McKee, C. F. \& Klein, R. I. 2004, ApJ, 611, 399

%\bibitem{14}
Krumholz, M. R., Klein, R. I.  \& McKee, C. F. 2007, ApJ, 656, 959

%\bibitem{15}
Kunz, M. W. \& Mouschovias, T. Ch. 2009, MNRAS, 399, L94

%\bibitem{16}
Lada, C. J. \& Lada, E. A. 2003, ARAA, 41, 57

%\bibitem{17}
Larson, R. 1985, MNRAS, 214, 379

%\bibitem{18}
Larson, R. 2003, in Galactic Star Formation Across the Stellar Mass
Spectrum, ASP Conf, Series, Vol. 297, eds. J. M. De Buizer and
N. S. van der Bliek, p65

%\bibitem[]{}
%Masunaga, H., Miyama, S. M. \& Inutsuka, S. I. 1998, ApJ, 495, 346

%\bibitem{19}
Matzner, C. D. \& McKee, C. F. 2000, ApJ, 545, 364

%\bibitem{20}
McKee, C. F. 1999, in in The Origin of Stars and Planetary Systems,
eds. C. J. Lada \& N. D. Kylafis (Kluwer), p29

%\bibitem{21}
McKee, C. F. \& Ostriker, E. C. 2007, ARAA, 45, 565

%\bibitem{22}
Mundy, L., Looney, L. \& Welch, W. J. 2000, in Protostars and Planets
IV, eds. V. Mannings, A. Boss and S. Russell (Arizona Unversity
Press), p355

%\bibitem{23}
Myers, P. C. 2008, ApJ, 687, 340

%\bibitem{24}
Nakamura, F. \& Li, Z.-Y. 2005, ApJ, 631, 411

%\bibitem{25}
Nakamura, F. \& Li, Z.-Y. 2007, ApJ, 662, 395

%\bibitem{26}
Offner, S., Klein, R. I., McKee, C. F. \& Krumholz, M. R. 2009, ApJ,
703, 131

%\bibitem{27}
Padoan, P. \& Nordlund, A. 2002, ApJ, 576, 870

%\bibitem{28}
Price, D. J. \& Bate, M. R. 2009, MNRAS, 398, 33

%\bibitem{29}
Shu, F. H., Adams, F. C., \& Lizano, S. 1987,
\araa, 25, 23

%\bibitem{30}
Shu, F. H. et al. 1999, in The Origin of Stars and Planetary Systems,
eds. C. J. Lada \& N. D. Kylafis (Kluwer), p193

%\bibitem{31}
Shu, F. H., Li, Z.-Y. \& Allen, A. 2004, ApJ, 601, 930

%\bibitem{32}
Smith, R. J., Longmore, S. \& Bonnell, I. 2009, MNRAS, 400, 1775

%\bibitem{33}
Spitzer, L. 1978, Physical Processes in the Interstellar Medium (Willey)

%\bibitem{34}
Tilley, D. \& Pudritz, R. 2007, ApJ, MNRAS, 382, 73

%\bibitem[]{}
%Truelove, J. K., Klein, R. I. et al. 1998, ApJ, 495, 821 

Urban, A., Martel, H. \& Evans, N. J. 2009, ApJ, in press

Vaillancourt, J. E. et al. 2008, ApJ, 679, L25

Wang, P., Li, Z.-Y., Abel, T. \& Nakamura, F. 2010, ApJ, 709, 27

Zinnecker, H. 1982, NYASA, 395, 226

\end{document}